\RequirePackage{fix-cm}
\documentclass[smallextended]{svjour3}       
\smartqed  
\usepackage{graphicx}
\usepackage[utf8]{inputenc}
\usepackage{amsmath}
\usepackage[nodisplayskipstretch]{setspace}
\usepackage{authblk}
\usepackage{hyperref}
\usepackage[none]{hyphenat}
\usepackage{float}
\begin{document}

\title{Efficient Deterministic Secure Quantum Communication protocols using multipartite entangled states
}


\author{Dintomon Joy         \and
        Supin P Surendran      \and
        Sabir M 
}


\institute{Dintomon Joy \and Supin P Surendran \and Sabir M \at
              Department of Physics, Cochin University of Science and Technology, Kochi - 682 022, India \\
              \email{dintomonjoy@cusat.ac.in}           
               }

\date{Received: date / Accepted: date}

\maketitle

\begin{abstract}
\sloppy We propose two deterministic secure quantum communication (DSQC) protocols employing three-qubit GHZ-like states
and five-qubit Brown states as quantum channels for secure transmission of information in units of two bits and three
bits using multipartite teleportation schemes developed here. In these schemes, the sender's capability in selecting
quantum channels and the measuring bases leads to improved qubit efficiency of the protocols.

\end{abstract}

\section{Introduction}
\label{intro}

\sloppy The field of quantum cryptography which offers  unconditional security in communication between legitimate users
has emerged as  an important area of research in this information age. In 1984, Bennett and Brassard\cite{Bennett}
proposed the first Quantum Key Distribution (QKD) protocol for secure communication. Following this, several
 protocols\cite{Ekert}\cite{Long}were suggested all of which required a unique secret key to be shared
 between the users before communication. Later developments showed how secure quantum communication could be achieved
 without the need for shared secret keys. Among the two group of protocols, the  deterministic secure quantum
 communication (DSQC) protocols\cite{Lo} the receiver can read out the message only after the exchange of
 at least one bit of classical information per qubit. But, in  quantum secure direct communication(QSDC) the message is read out
directly. Compared with QSDC, the DSQC protocols are more secure in that the message-carrying qubits need not be transmitted
through external channels. Both protocols, however, require classical communication during the error checking and eavesdropping
detection  processes.

In 1999, Shimizu and Imoto\cite{Shimizu} proposed the first DSQC protocol using EPR pairs and Bell measurements.
Later, Beige \textit{et.al.}\cite{Beige} proposed a scheme based on single photons. In 2004, Yan and Zhang\cite{Yan}
found the first secure DSQC protocol based on teleportation. Following this work, several
schemes employing multipartite entangled channels and teleportation were proposed\cite{HJCao}\cite{DXG08}\cite{XGC09}\cite{QCY13}\cite{Gaot}\cite{Cao}\cite{Zhang}
 Many DSQC protocols  based on  entanglement swapping\cite{Gaoswap}\cite{Man}\cite{Xiu}\cite{Gaoepr}\cite{ManZ}
 and order rearrangement of particles\cite{Zhu}\cite{LiXH} were also suggested.
The advantage of teleportation based protocols is that these are more secure even in noisy channel\cite{Hassan} and are
suitable for quantum error correction\cite{Li}.

In this paper we propose two new  DSQC protocols one of these using the three-qubit GHZ-like state and the other
using the five-qubit Brown states for secure transmission of information in units two bits
and three bits employing multipartite teleportation techniques. In these schemes, it is   the sender who decides which entanglement
channel and measurement basis are to be used. The receiver and  the possible eavesdroppers are ignorant of the choices.
The receiver  performs unitary operations based on the classical information obtained from the sender and finally
performs joint measurements to read out the secret message.

This paper is organized as follows. In section ~\ref{sec2}, we present the teleportation scheme of a special class of
two particle state using GHZ-like state and a special class of three particle state using five-qubit Brown state in \ref{sec3}. The details of our new DSQC protocols using GHZ-like state and Brown state are given in \ref{sec3}.
 In section \ref{sec4}, we discuss the
possible eavesdropping attacks. In section \ref{sec6}, we discuss and compare the efficiency of our protocol with other existing
protocols. The final section contains our conclusions.

\section{Teleportation of  special classes of two and three particle states}
\label{sec2}

\subsection{Teleportation of a two particle state using GHZ-like state}
\label{sec2.1}
\sloppy In 2014, Nandi and Mazumdar\cite{Nandhi} proposed a scheme for teleportation of a special form of  two particle
state, $\left|\tau \right\rangle_{12} = \alpha (\left|00\right\rangle + \left|11\right\rangle)_{12} +
\beta(\left|01\right\rangle + \left|10\right\rangle)_{12}$
using GHZ-like state
\begin{equation}
\label{eqn a}
\left|\Phi_{G} \right\rangle_{345}=\frac{1}{\sqrt{2}}\{\left|0\right\rangle_3
\left|\psi^+\right\rangle + \left|1\right\rangle_3 \left|\phi^+\right\rangle_{45}\}
\end{equation}
as entanglement channel.
\sloppy Here, the states $\left|\phi^{\pm}\right\rangle=\frac{1}{\sqrt{2}}\{\left|01\right\rangle\pm\left|10\right\rangle\}$ and
$\left|\psi^{\pm}\right\rangle=\frac{1}{\sqrt{2}}\{\left|00\right\rangle\pm\left|11\right\rangle\}$ represent
the Bell basis states. A three particle joint measurement is performed in the measurement basis:
$\left|\zeta^{\pm} \right\rangle= \frac{1}{2}\{(\left|001\right\rangle  \left|111\right\rangle)
\pm (\left|010\right\rangle - \left|100\right\rangle)\}$ and $\left|\eta^{\pm} \right\rangle= \frac{1}{2}
\{(\left|000\right\rangle - \left|110\right\rangle) \pm (\left|011\right\rangle - \left|101\right\rangle)\}$, to achieve
this teleportation. In the DSQC protocol proposed in section:\ref{sec3}, we need a  teleportation scheme
for a related two particle state.
\begin{equation}
\left|\tau' \right\rangle_{12}= \alpha (\left|00\right\rangle - \left|11\right\rangle)_{12} + \beta(\left|01\right\rangle -
\left|10\right\rangle)_{12}\\
\end{equation}
We show, here, that this can be done by utilizing the Nandi-Mazumdar  method by choosing a different GHZ-like state
\begin{eqnarray}
\label{eqn b}
\left|\Phi'_{G} \right\rangle_{345}=\frac{1}{\sqrt{2}}\{\left|0\right\rangle_3 \left|\phi^-\right\rangle +
\left|1\right\rangle_3 \left|\psi^-\right\rangle_{45}\}\\ \nonumber
\end{eqnarray}
as the entanglement channel. Then the total wave function is  $\Gamma_{12345}=\left|\tau' \right\rangle_{12} \otimes
\left|\Phi'_{G} \right\rangle_{345}$ and  Alice performs a three particle joint measurement on particles in
her possession $(1,2,3)$ in a new measurement basis: $\left|\zeta'^{\pm} \right\rangle=
\frac{1}{2}\{(\left|001\right\rangle - \left|111\right\rangle)   \pm (\left|010\right\rangle -
\left|100\right\rangle)\}$ and $\left|\eta'^{\pm} \right\rangle= \frac{1}{2}
\{(\left|000\right\rangle - \left|110\right\rangle) \pm (\left|011\right\rangle - \left|101\right\rangle)\}$.
After getting the information about measurement result from Alice, Bob performs
unitary operations on his particles $(4,5)$ as given in  Table:\ref{tab} and recovers the teleported state.
It may be noted that the table:\ref{tab} coincides with the table given in \cite{Nandhi} except for the changes
in labels representing measurement result and the state of Bob's particles (4,5).
\begin{table}
\begin{tabular}{l l l l}
\hline\noalign{\smallskip}
Alice's     & Classical   & State of Bob's particles (4,5)& Unitary \\
measurement & information &    & operation\\
result      &             &             &\\
\noalign{\smallskip}\hline\noalign{\smallskip}
$\left|\zeta'^{+}\right\rangle$ & 00& $\alpha(\left|00\right\rangle-\left|11\right\rangle)+\beta(\left|01\right\rangle-\left|10\right\rangle)$ & $(I \otimes I)$ \\
$\left|\zeta'^{-}\right\rangle$ & 01 & $\alpha(\left|00\right\rangle-\left|11\right\rangle)-\beta(\left|01\right\rangle-\left|10\right\rangle)$ & $(\sigma_z \otimes \sigma_{z})$ \\
$\left|\eta'^{+}\right\rangle$ & 10 & $\alpha(\left|01\right\rangle-\left|10\right\rangle)+\beta(\left|00\right\rangle-\left|11\right\rangle)$ & $(I \otimes \sigma_{x})$\\
$\left|\eta'^{-}\right\rangle$ & 11 & $\alpha(\left|01\right\rangle-\left|10\right\rangle)-\beta(\left|00\right\rangle-\left|11\right\rangle)$ & $(\sigma_z \otimes i\sigma_{y})$ \\
\noalign{\smallskip}\hline
\end{tabular}
\caption{Classical information corresponding to Alice's measurement result, state of particles (4,5) and
unitary operations performed by Bob.}
\label{tab}
\end{table}

\subsection{Teleportation of a three particle state using five-qubit Brown state}
\label{sec2.2}
We extend the two-qubit teleportation scheme given above for the teleportation of a three particle state
\begin{eqnarray}
\label{3state}
\left|\psi\right\rangle_{123} &=& \{\alpha(\left|000\right\rangle-\left|111\right\rangle) + \beta(\left|001\right\rangle+\left|110\right\rangle)\\ \nonumber
&+&\gamma(\left|010\right\rangle+\left|101\right\rangle)+ \delta(\left|011\right\rangle-\left|100\right\rangle)\}_{123}\\ \nonumber
\end{eqnarray}
where $\alpha, \beta, \gamma$ and $\delta$ are  unknown coefficients satisfying the relation
 $|\alpha|^2+ |\beta|^2+ |\gamma|^2 +|\delta|^2 = 1$.

The quantum channel is chosen as the maximally entangled five-qubit Brown state\cite{Brown}.
Brown \textit{et.al.,} initially obtained this state via a numerical optimization procedure and later
Muralidharan and Panigrahi\cite{Murali} proposed a simple theoretical
   method for the physical realization of this state.
 The five-qubit Brown state is given by

\begin{multline}
\label{check1}
\left|\Psi\right\rangle_{45678} = \frac{1}{2} \{ \left|001\right\rangle\left|\phi^{-}\right\rangle + \left|010\right\rangle\left|\psi^{-}\right \rangle + \left|100\right\rangle\left|\phi^{+}\right\rangle + \left|111\right\rangle\left|\psi^{+}\right\rangle\}_{45678}
\end{multline}
This can be  rewritten in equivalent form  in the following order as given in \cite{Lin}.
\begin{equation}
\label{5state}
\left|\Psi\right\rangle_{67\:458} =\frac{1}{2}\{\left|00\right\rangle\left|G_{010}\right\rangle-\left|01\right\rangle
\left|G_{111}\right\rangle+\left|10\right\rangle\left|G_{001}\right\rangle-\left|11\right\rangle\left|G_{100}
\right\rangle\}_{67\:458}\\
\end{equation}

Here, the three-qubit states defined by
\begin{equation}
 \left|G_{ijk}\right\rangle= \frac{1}{\sqrt{2}}\{\left|0\right\rangle \left|j\right\rangle \left|k\right\rangle +
  (-1)^i \left|1\right\rangle\left|j \oplus 1\right\rangle\left|k \oplus 1\right\rangle\}
  \end{equation}
where $i, j, k $  takes values $ \{0,1\}$ and $\oplus$ represents  addition modulo 2,  are the GHZ\cite{GHZ} states.
 The equation:\ref{3state} can be expressed in terms of these states as $$\left|\psi\right\rangle_{123}=
 \sqrt{2}(\alpha\left|G_{100}\right\rangle + \beta\left|G_{001}\right\rangle +\gamma\left|G_{010}\right\rangle+
 \delta\left|G_{111}\right\rangle)_{123}$$
Alice, now, performs a five-particle joint measurement on her particles $(1,2,3,6,7)$ in the  measurement basis
 $\{\left|\Phi_{1,2...16}\right\rangle \}$ given below:
\begin{eqnarray}
\label{5basis}
\left|\Phi_{1,2}\right\rangle= \frac{1}{2}\{\left|G_{010}\right\rangle\left|00\right\rangle-\left|G_{111}\right\rangle\left|01\right\rangle\pm\left|G_{001}\right\rangle\left|10\right\rangle\mp\left|G_{100}\right\rangle\left|11\right\rangle\}\\ \nonumber
\left|\Phi_{3,4}\right\rangle= \frac{1}{2}\{\left|G_{010}\right\rangle\left|00\right\rangle+\left|G_{111}\right\rangle\left|01\right\rangle\pm\left|G_{001}\right\rangle\left|10\right\rangle\pm\left|G_{100}\right\rangle\left|11\right\rangle\}\\ \nonumber
\left|\Phi_{5,6}\right\rangle= \frac{1}{2}\{\left|G_{010}\right\rangle\left|01\right\rangle-\left|G_{111}\right\rangle\left|00\right\rangle\pm\left|G_{001}\right\rangle\left|11\right\rangle\mp\left|G_{100}\right\rangle\left|10\right\rangle\}\\ \nonumber
\left|\Phi_{7,8}\right\rangle= \frac{1}{2}\{\left|G_{010}\right\rangle\left|01\right\rangle+\left|G_{111}\right\rangle\left|00\right\rangle\pm\left|G_{001}\right\rangle\left|11\right\rangle\pm\left|G_{100}\right\rangle\left|10\right\rangle\}\\ \nonumber
\left|\Phi_{9,10}\right\rangle= \frac{1}{2}\{\left|G_{010}\right\rangle\left|10\right\rangle-\left|G_{111}\right\rangle\left|11\right\rangle\pm\left|G_{001}\right\rangle\left|00\right\rangle\mp\left|G_{100}\right\rangle\left|01\right\rangle\}\\ \nonumber
\left|\Phi_{11,12}\right\rangle= \frac{1}{2}\{\left|G_{010}\right\rangle\left|10\right\rangle+\left|G_{111}\right\rangle\left|11\right\rangle\pm\left|G_{001}\right\rangle\left|00\right\rangle\pm\left|G_{100}\right\rangle\left|01\right\rangle\}\\ \nonumber
\left|\Phi_{13,14}\right\rangle= \frac{1}{2}\{\left|G_{010}\right\rangle\left|11\right\rangle-\left|G_{111}\right\rangle\left|10\right\rangle\pm\left|G_{001}\right\rangle\left|01\right\rangle\mp\left|G_{100}\right\rangle\left|00\right\rangle\}\\ \nonumber
\left|\Phi_{15,16}\right\rangle= \frac{1}{2}\{\left|G_{010}\right\rangle\left|11\right\rangle+\left|G_{111}\right\rangle\left|10\right\rangle\pm\left|G_{001}\right\rangle\left|01\right\rangle\pm\left|G_{100}\right\rangle\left|00\right\rangle\}\\ \nonumber
\end{eqnarray}
Alice then communicates her measurement result to Bob via four bits of classical information. In order to recover the
teleported state, appropriate unitary operations are performed by Bob on his particles $(4,5,8)$ as given in
 table:~\ref{tab: y}.
\begin{table}
\centering
\begin{tabular*}{\textwidth}{c @{\extracolsep{\fill}} ccccc}
\hline\noalign{\smallskip}
Alice's result & State of qubits (6,7,8) & Unitary \\
and Classical& &operation\\
 information & &\\
\noalign{\smallskip}\hline\noalign{\smallskip}
$\left|\Phi_{1}\right\rangle 0000$  &$\sqrt{2}(\alpha\left|G_{100}\right\rangle + \beta\left|G_{001}\right\rangle +\gamma\left|G_{010}\right\rangle+ \delta\left|G_{111}\right\rangle)$ & $ I \otimes  I \otimes I $\\
$\left|\Phi_{2}\right\rangle 0001$ &$\sqrt{2}(-\alpha\left|G_{100}\right\rangle -\beta\left|G_{001}\right\rangle +\gamma\left|G_{010}\right\rangle+ \delta\left|G_{111}\right\rangle)$ & $ \sigma_z\otimes \sigma_z \otimes I $\\
$\left|\Phi_{3}\right\rangle 0010$  &$\sqrt{2}(-\alpha\left|G_{100}\right\rangle + \beta\left|G_{001}\right\rangle +\gamma\left|G_{010}\right\rangle- \delta\left|G_{111}\right\rangle)$ & $ \sigma_z \otimes I \otimes \sigma_z$\\
$\left|\Phi_{4}\right\rangle 0011$ &$\sqrt{2}(\alpha\left|G_{100}\right\rangle - \beta\left|G_{001}\right\rangle +\gamma\left|G_{010}\right\rangle- \delta\left|G_{111}\right\rangle)$ & $ I\otimes \sigma_z \otimes \sigma_z $\\
$\left|\Phi_{5}\right\rangle 0100$  &$\sqrt{2}(-\alpha\left|G_{001}\right\rangle - \beta\left|G_{100}\right\rangle -\gamma\left|G_{111}\right\rangle- \delta\left|G_{010}\right\rangle)$ & $ I\otimes I \otimes \sigma_x $\\
$\left|\Phi_{6}\right\rangle 0101$  &$\sqrt{2}(\alpha\left|G_{001}\right\rangle + \beta\left|G_{100}\right\rangle -\gamma\left|G_{111}\right\rangle- \delta\left|G_{010}\right\rangle)$ & $ \sigma_z \otimes \sigma_z  \otimes\sigma_x  $\\
$\left|\Phi_{7}\right\rangle 0110$  &$\sqrt{2}(\alpha\left|G_{001}\right\rangle - \beta\left|G_{100}\right\rangle -\gamma\left|G_{111}\right\rangle+ \delta\left|G_{010}\right\rangle)$ & $ \sigma_z\otimes I \otimes i\sigma_{y}$\\
$\left|\Phi_{8}\right\rangle 0111$ &$\sqrt{2}(-\alpha\left|G_{001}\right\rangle + \beta\left|G_{100}\right\rangle -\gamma\left|G_{111}\right\rangle+ \delta\left|G_{010}\right\rangle)$ & $ I \otimes \sigma_z \otimes i\sigma_{y}$\\
$\left|\Phi_{9}\right\rangle 1000$  &$\sqrt{2}(\alpha\left|G_{111}\right\rangle + \beta\left|G_{010}\right\rangle +\gamma\left|G_{001}\right\rangle+ \delta\left|G_{100}\right\rangle)$ & $ I \otimes \sigma_x \otimes I$\\
$\left|\Phi_{10}\right\rangle 1001$  &$\sqrt{2}(-\alpha\left|G_{111}\right\rangle - \beta\left|G_{010}\right\rangle+\gamma\left|G_{001}\right\rangle+ \delta\left|G_{100}\right\rangle)$ & $ \sigma_z \otimes i\sigma_{y} \otimes I $\\
$\left|\Phi_{11}\right\rangle 1010$  &$\sqrt{2}(-\alpha\left|G_{111}\right\rangle + \beta\left|G_{010}\right\rangle +\gamma\left|G_{001}\right\rangle- \delta\left|G_{100}\right\rangle)$ & $ \sigma_z\otimes\sigma_x  \otimes\sigma_z $\\
$\left|\Phi_{12}\right\rangle 1011$  &$\sqrt{2}(\alpha\left|G_{111}\right\rangle - \beta\left|G_{010}\right\rangle +\gamma\left|G_{001}\right\rangle- \delta\left|G_{100}\right\rangle)$ & $I \otimes i\sigma_{y} \otimes \sigma_z $\\
$\left|\Phi_{13}\right\rangle 1100$  &$\sqrt{2}(-\alpha\left|G_{010}\right\rangle - \beta\left|G_{111}\right\rangle -\gamma\left|G_{100}\right\rangle- \delta\left|G_{001}\right\rangle)$ & $ \sigma_x\otimes I \otimes I $\\
$\left|\Phi_{14}\right\rangle 1101$  &$\sqrt{2}(\alpha\left|G_{010}\right\rangle + \beta\left|G_{111}\right\rangle -\gamma\left|G_{100}\right\rangle- \delta\left|G_{001}\right\rangle)$ & $ i\sigma_{y}\otimes \sigma_z \otimes I $\\
$\left|\Phi_{15}\right\rangle 1110$  &$\sqrt{2}(\alpha\left|G_{010}\right\rangle - \beta\left|G_{111}\right\rangle -\gamma\left|G_{100}\right\rangle+ \delta\left|G_{001}\right\rangle)$ & $ i\sigma_{y}\otimes I \otimes \sigma_z$\\
$\left|\Phi_{16}\right\rangle 1111$  &$\sqrt{2}(-\alpha\left|G_{010}\right\rangle + \beta\left|G_{111}\right\rangle -\gamma\left|G_{100}\right\rangle+ \delta\left|G_{001}\right\rangle)$ & $ \sigma_x\otimes \sigma_z \otimes \sigma_z$\\
\noalign{\smallskip}\hline
\end{tabular*}
\caption{Alice's measurement results and corresponding operations performed by Bob}
\label{tab: y}
\end{table}

In a similar fashion another three-qubit state $\left|\psi'\right\rangle_{123} =
\{\alpha(\left|000\right\rangle+\left|111\right\rangle) +
\beta(\left|001\right\rangle-\left|110\right\rangle)+\gamma(\left|010\right\rangle-\left|101\right\rangle)+
\delta(\left|011\right\rangle+\left|100\right\rangle)\}_{123}$ needed in the  DSQC protocol proposed in section:\ref{sec3}
can be teleported. The appropriate five-qubit channel
  $\left|\Psi'\right\rangle $ and the measurement basis $\left|\Phi'_{1,2...16}\right\rangle$ are obtained
   by making the replacements ($\left|G_{100}\right\rangle \rightarrow \left|G_{000}\right\rangle$, $\left|G_{001}\right\rangle
   \rightarrow\left|G_{101}\right\rangle$, $\left|G_{010}\right\rangle \rightarrow \left|G_{110}\right\rangle$ and
  $\left|G_{111}\right\rangle \rightarrow    \left|G_{011}\right\rangle$) in equations:~\ref{5state} and ~\ref{5basis} respectively.
    \textit{i.e.,} $\left|\Psi\right\rangle_{67458}$ changes to $\left|\Psi'\right\rangle_{67458}$ and
     $\left|\Phi_{1,2...16}\right\rangle$ changes to $\left|\Phi'_{1,2...16}\right\rangle$. The recovery operations
to be performed by Bob, corresponding to Alice's classical information, are same as in table:\ref{tab: y}with the labels
   representing measurement results and state of Bob's particles  changed accordingly.
\section{New Deterministic Secure Quantum Communication protocols}
Applying the teleportation schemes in the preceding section, we propose two novel DSQC protocols for communication of
information in units of 2 bits or 3 bits.
\label{sec3}
\subsection{DSQC protocol using GHZ-like state}
\label{sec3.1}
 Our protocol for sending information in units of 2 bits makes use of the GHZ-like state as the quantum channel
 and  involves  the following steps.

\textbf{1. Preparation of GHZ-like states:} The sender, Alice, prepares copies of two orthogonal GHZ-like states
$\left|\Phi_{G}\right\rangle$ and $\left|\Phi'_{G} \right\rangle$, their number depending on the size of secret messages to be sent.
 The ordered sequence of three particle GHZ-like state $\left|\Phi_{G}\right\rangle_{345}$ is given by $[(P_{3}^{1}, P_{4}^{1},
  P_{5}^{1}), (P_{3}^{2}, P_{4}^{2}, P_{5}^{2}),.... (P_{3}^{n}, P_{4}^{n}, P_{5}^{n})]$
and the sequence of $\left|\Phi'_{G}\right\rangle_{345}$ is given by $[(P_{3}^{'1}, P_{4}^{'1}, P_{5}^{'1}), (P_{3}^{'2},
 P_{4}^{'2}, P_{5}^{'2}),.... (P_{3}^{'n}, P_{4}^{'n}, P_{5}^{'n})]$. The entangled states in each of these sets are given a unique number, with the superscripts representing  the order in the sequence and the subscript representing particle labels.\\

\textbf{2. Splitting Home Block and Travel Block:} The block of all the first particles in the sequences
$\left|\Phi_{G}\right\rangle$ and $\left|\Phi'_{G} \right\rangle$, $[P_{3}^{1}, P_{3}^{'1}, P_{3}^{2},
 P_{3}^{'2},... P_{3}^{n}, P_{3}^{'n}]$ are kept at sender's location and we refer to them as the home block.
 The block of all second and third
particles from both sequences  $[(P_{4}^{1}, P_{5}^{1}), (P_{4}^{2}, P_{5}^{2}),....,
(P_{4}^{n},P_{5}^{n})],$ and  $[(P_{4}^{'1}, P_{5}^{'1}),(P_{4}^{'2}, P_{5}^{'2})$
 $....,(P_{4}^{'n}, P_{5}^{'n})]$ are referred to as the travel block. \\

\textbf{3. Arrangement of particles for Error checking and Message transmission:}
Alice selects $ \frac{N}{2}$ channels
 for error checking, where N represents the even number of secret information to be transmitted. The order and position
 of corresponding particles in the home block are noted by Alice and the leftover particles gets reserved for message
  transmission. In order to send information bits  00 or 01 Alice uses
 the channel $\left|\Phi_{G}\right\rangle$ and for sending 10 or 11 she uses $\left|\Phi'_{G}\right\rangle$ . Alice arranges
 all the particles in both travel and home block according to the particular message to be sent.  Alice sends the
 travel block to Bob and retains the home block with herself.\\

\textbf{4. Error checking:} On receiving the travel block, Bob sends a signal back to Alice confirming the reception.
  Alice, then, initiates the error checking process by performing single particle measurement on her particles initially
  chosen for error checking. After the measurement, Alice announces the positions of particles that were chosen for
  error checking. Bob performs either Bell measurement or single particle measurement on the corresponding particles in
  his possession according to the information received  from Alice and he, in turn, announces his measurement outcomes.
  Alice, now, checks for correlations in the measurement results by comparing it with equations: \ref{eqn a} and \ref{eqn b}. If the error rate is high
  they abort the communication. Otherwise they proceed to the next step.

\textbf{5. Teleportation:} In the scheme of communicating in units of two bit the users make an agreement
to encode the bits $00, 01, 10$ and $11$ in the Bell states $\left|\psi^{+}\right\rangle$,
$\left|\phi^{+}\right\rangle$, $\left|\psi^{-}\right\rangle$, and $\left|\phi^{-}\right\rangle$ respectively.
With appropriate choice the  coefficients $\alpha$ and $\beta$  in
 $\left|\tau \right\rangle_{12}$ and $\left|\tau'\right\rangle_{12}$ (effectively reducing these to the Bell states)
 information can be conveyed securely using the teleportation scheme described earlier. For instance, if Alice wants to
   send 10 or 11, she choose the state $\left|\tau' \right\rangle_{12}$ and channel $\left|\Phi'_{G} \right\rangle_{345}$.
   The coefficient $\alpha$ or $ \beta$ is made zero depending on  whether the information is 11 or 10.
 Alice, then, follows the teleportation procedure given in section:\ref{sec2}. Finally, Bob performs
    a Bell measurement to read out the secret information.

Our protocol works efficiently when the sender prepares the secret message beforehand. For example, consider the
 case where Alice wants to communicate a four bit secret information `0011' to Bob. She prepares two channels,
  $\left|\Phi_{G} \right\rangle$, $\left|\Phi'_{G} \right\rangle$ and arrange them in the same order as that of
 the messages \textit{i.e.,} $\left|\Phi_{G} \right\rangle$ taken first and $\left|\Phi'_{G} \right\rangle$ second.
 The steps 2, 3 and 4 are followed. Finally, the state $\left|\tau\right\rangle_{12}$ with coefficient $\beta=0$ or
  $\left|\psi^{+}\right\rangle$ and  $\left|\tau'\right\rangle_{12}$ with $\alpha=0$ or $\left|\phi^{-}\right\rangle$,
 corresponding to information 00 and 11, are teleported using the appropriate schemes given in section:\ref{sec2}.
In this scheme, it may be noted that the participants (receiver and eavesdropper) are not aware of the different entanglement channels
   and measurement basis employed by the sender. The receiver performs unitary operations based on the classical
 information obtained from sender and read out the secret messages by performing Bell measurements.

\subsection{DSQC protocol using five-qubit maximally entangled Brown states.}
\label{5dsqc}
By a straight foreword generalization of the procedure given above, Alice can transmit information in units of 3 bits
by using the five-qubit Brown states as quantum channels and employing the teleportation scheme given in section:\ref{sec2.2}. Before
transmitting messages Alice and Bob agree that the states $\left|G_{ijk}\right\rangle$, where ${i,j,k}$ take
values $\{0,1\}$ correspond to 8 different messages $(000, 001, 010, 011, 100, 101, 110, 111)$ respectively.
For instance, if Alice wants to transmit a six bit secret information `100 110' to Bob, she chooses two
channels $\left|\Psi\right\rangle$ and $\left|\Psi'\right\rangle$ and then arrange them in the order of messages
 \textit{i.e.,} $\left|\Psi\right\rangle$ taken first and $\left|\Psi'\right\rangle$ second. $\left|G_{100}\right\rangle$
 and $\left|G_{110}\right\rangle$ are the corresponding messages and they are prepared by setting the coefficient $\alpha=1$,
 other coefficients to zero in equation:\ref{3state} and $\gamma=1$, other coefficients to zero in its conjugate equation
  $\left|\psi'\right\rangle$ respectively. Alice follows the steps discussed in the previous section and finally
 teleport the states as given in section:\ref{sec2}. Here, 8 bits of classical information is used up in this process.

\section{Security Analysis}
\label{sec4}
In teleportation-based DSQC protocols, the information carrying qubits do not travel and thus the security of
information solely depends on the security of quantum channel shared between Alice and Bob. The most common attack
strategies of an eavesdropper for DSQC using GHZ-like is discussed below.

\textbf{Measure and resend Attack:}
In this type of attack, the eavesdropper Eve may capture the travel block from Alice and measure all the qubits
 in either Bell basis or perform single particle measurement on them and resend it to Bob. After this process,
  neither Alice-Bob pair nor any Alice-Eve pair shares entanglement. Therefore it is impossible for Eve to
  gain any useful information from Alice's side using this type of attack, as long as she employs teleportation
   procedure to transmit the secret information. Further, in the first error checking process the correlation results
    itself will reveal the presence of Eve and the communication is abandoned.

\textbf{Intercept and resend Attack:}
The eavesdropper may try to intercept the travel block and replace it with particles (6, 7) which is a part of random
 GHZ-like states prepared by Eve and resend them to Bob. Now, Eve establishes an entanglement relation with Alice's
  particle and may receive the teleported message. But, on checking the correlations given in equations:~\ref{eqn a}
   and ~\ref{eqn b}, Alice detects the presence of Eve. If the error rate is high, she informs Bob to stop the entire
    process.

\textbf{Entanglement Attack: }
In this attack\cite{Yan}\cite{Gaot}, the eavesdropper may capture the travel block and entangle the ancilla particles
 (6,7) in the state $\left|00\right\rangle_{67}^n$ with the $n^{th}$ order of particles (4,5) in the travel block, where
  $n$ represents the order of entangled GHZ-like particles in the travel block. Eve uses CNOT gate to entangle ancilla
  qubits with travel block. The particles (4,5) of GHZ-like state acts as control bits and (6,7) as target bits
  respectively. This changes the total state to
\begin{eqnarray}
\label{eqn c}
\left|\Psi\right\rangle_{34567}&=& CNOT(5;7) CNOT(4;6)\left|\Phi_G\right\rangle_{345} \otimes \left|00\right\rangle_{67}\\\nonumber
&=& \frac{1}{2}[\left|0\right\rangle_3 (\left|\phi^{+}\right\rangle_{45} \left|\phi^{+}\right\rangle_{67} + \left|\phi^{-}
\right\rangle_{45} \left|\phi^{-}\right\rangle_{67})\\\nonumber
 &+& \left|1\right\rangle_3 (\left|\psi^{+}\right\rangle_{45} \left|\psi^{+}\right\rangle_{67} + \left|\psi^{-}
 \right\rangle_{45} \left|\psi^{-}\right\rangle_{67})]\\\nonumber
\end{eqnarray}
The same operation changes the state of $\left|\Phi'_G\right\rangle_{345}$ to
\begin{eqnarray}
\label{eqn d}
\left|\Psi'\right\rangle_{34567}&=& \frac{1}{2}[\left|0\right\rangle_3 (\left|\phi^{+}\right\rangle_{45} \left|
\phi^{-}\right\rangle_{67} + \left|\phi^{-}\right\rangle_{45} \left|\phi^{+}\right\rangle_{67})\\  \nonumber
&+& \left|1\right\rangle_3 (\left|\psi^{+}\right\rangle_{45} \left|\psi^{-}\right\rangle_{67} + \left|\psi^{-}
\right\rangle_{45} \left|\psi^{+}\right\rangle_{67})]\\\nonumber
\end{eqnarray}

Equation:~\ref{eqn c} shows, the state of (4,5) and (6,7) are correlated and equation:~\ref{eqn d} shows that they
are anti-correlated.
By entangling ancillary particles to travel block Eve gets access to Alice's information carrying qubits. But,
 during the error checking process, Alice can detect the presence of Eve 50\% of times by checking relations given
  in equations ~\ref{eqn a} and~\ref{eqn b}. Alice informs Bob to continue the process if the error rate is less than
   50\% and otherwise to abandon. In the case where Eve goes undetected, the probability of secret message getting
    leaked is still $50\%$ due to the superposition of states existing between particles $(4,5,6,7)$.

\section{Efficiency Analysis}
\label{sec6}
In 2000, Cabello\cite{Cabello} put forward a simple definition to find the qubit efficiency of a given protocol.
 Here the qubit efficiency $\eta_2$ depends on the ratio of the number of messages transmitted $b_s,$ to the sum
  of total number of classical bits $b_t$ utilized to decode the message and the total number of  qubits $q_t$
  used in the protocol. Note that, here the decoy qubits and the classical bits used for eavesdrop checking are
  not included to calculate the efficiency.
\begin{equation}
\eta_2= \frac{b_s}{q_t+b_t}
\end{equation}
In 2012 Banerjee and Pathak\cite{AA} proposed a modified  measure of efficiency which considers the decoy qubits used
for eavesdrop checking. Since the number of  classical bits used in eavesdrop checking varies linearly with number
of decoy qubits used, inclusion of the number of decoy qubits in efficiency calculations seems more appropriate.
Using the definition of \cite{AA}(including decoy qubits) and the one by \cite{Cabello} (excluding decoy qubits),
we have calculated the qubit efficiency of different teleportation based protocols
 \cite{Yan}\cite{HJCao}\cite{DXG08}\cite{XGC09}\cite{QCY13} and have tabulated them in table:~\ref{eff}. We
find that the qubit efficiency of these protocols are $\leq 20\%$.
In these calculations we have taken the number of decoy states as equal to the number of transmitted qubits
following  suggestion in \cite{AA}

In our protocol with GHZ-like state as quantum channel, to communicate a four bit message $(b_s =4)$ we require
 two GHZ-like channels or 6 qubits \textit{i.e.,}$(q_t=6)$ and four bits of classical information $(b_t=4)$.
 As the classical signal from Bob after the receipt of travel block becomes asymptotically(N large) insignificant,
  it is not included in $b_t$. By definition of \cite{Cabello}, this leads to a qubit efficiency of $\eta$ = 40\%.
  But, in order to ensure security, we used $\frac{N}{2}$ GHZ-like channels for eavesdrop checking in the protocol.
    In our example, we used two GHZ-like states for eavesdrop checking. Hence the total number of qubits consumed
    becomes $(q_t=12)$. The efficiency calculated according to \cite{AA} gives $\eta$ = 25\%. In the case of
    sending a six bit message $(b_s =6)$ given in section:\ref{5dsqc}, using two five qubit channels $(q_t=10)$ and
    8 bits of classical information $(b_t=8)$ the qubit efficiency is $\eta$ = 33.33\% according to \cite{Cabello}.
     Using the definition of \cite{AA}, it becomes 21.43\%.
\begin{table}
\begin{tabular}{l l l l l l}
\hline\noalign{\smallskip}
DSQC      &  $ \eta_2 =$ Efficiency(\%) &   & Entanglement \\
protocol& Without decoy & With decoy  \cite{AA}& Channel  \\
& qubits \cite{Cabello}& qubits&\\
\noalign{\smallskip}\hline\noalign{\smallskip}
YZ04 \cite{Yan}&25.00 & 16.67 & Bell pairs\\
CS06 \cite{HJCao}& 16.67&09.52 & W-State\\
DXG08 \cite{DXG08}& 20.00&12.50 & W-State\\
XGC09 \cite{XGC09}& 30.00&18.75 & Six-particle\\
QCY13 \cite{QCY13}& 25.00&16.67 & Four-qubit Cluster \\
PP-1 &40.00 &25.00 & GHZ-like State\\
PP-2 &33.33 &21.43 & Five-Qubit Brown\\
\noalign{\smallskip}\hline
\end{tabular}
\caption{Comparing efficiency of Teleportation-based DSQC protocols. PP- Proposed protocol}
\label{eff}
\end{table}

\section{Conclusion}
\label{sec7}
 We have proposed an extension of Nandi-Mazumdar scheme for teleportation of a special class of two particle states
and have also developed a method for the teleportation of special class of three-particle state using five-qubit Brown state as entanglement channel. The different teleportation schemes introduced here are then combined effectively to build
 novel DSQC protocols. These works provide  new insight into the way in which teleportation can be used for efficient
  DSQC protocols. The comparison of qubit efficiency shows that the proposed protocol is better than the other DSQC
 protocols based on teleportation \cite{Yan}\cite{HJCao}\cite{DXG08}\cite{XGC09}\cite{QCY13}. This increase in efficiency
  results from the sender's capability of choosing different channels and measurement basis independently. Even though our
protocol requires the sender to prepare the secret message beforehand, the efficiency factor gives it an edge
over other existing protocols. We suggest that the efficiency of using five-qubit Brown state can be further
increased if the sender employs more entanglement channels and appropriate measurement basis.

\section{Acknowledgements}
We thank Kerala State Council for Science, Technology and Environment for providing the financial support for this work. The authors also thank Kishore Thapliyal and Rishi Dutt Sharma for their suggestions regarding efficiency analysis.

\end{document}